\documentclass[aps,pra,twocolumn,superscriptaddress,amssymb,showpacs]{revtex4-1}
\usepackage{graphicx}
\usepackage{epstopdf}

\begin{document}

\title{  Engineering of  oscillatory quantum  states by parametric excitation }

\author{T.~V.~Gevorgyan}
\email[]{t_gevorgyan@ysu.am}
\affiliation{Institute for Physical Researches, National Academy
of Sciences,\\Ashtarak-2, 0203, Ashtarak, Armenia}

\author{G.~Yu.~Kryuchkyan}
\email[]{kryuchkyan@ysu.am}
\affiliation{Institute for Physical Researches,
National Academy of Sciences,\\Ashtarak-2, 0203, Ashtarak,
Armenia}\affiliation{Yerevan State University, Alex Manoogian 1, 0025,
Yerevan, Armenia}

\begin{abstract}
We consider preparation of nonclassical oscillatory states in a
degenerate parametric oscillator combined with phase modulation.
In this scheme intracavity oscillatory mode is excited by train of
Gaussian laser pulses through degenerate down-conversion process
and phase modulation element inserted in a cavity leads to
anharmonicity of oscillatory mode.  We demonstrate production of
nonclassical oscillatory states with two-fold symmetry in
phase-space  including  the superposition of Fock states and
quantum localized states on the level of few excitation numbers
and in over-transient dissipative regime.

\end{abstract}



\maketitle

\section{Introduction}

There is currently a wide effort to construct various artificial
nonlinear oscillatory systems showing quantum behavior.  Such
systems operated in quantum regime have become more important in
both fundamental and applied sciences, particularly, for
implementation of basic quantum optical systems, in engineering of
nonclassical states and quantum logic. In these systems the
efficiency of quantum effects requires a high nonlinearity with
respect to dissipation. The nonlinearity breaks the equidistance
of oscillatory energy levels that allows selectively excites the
oscillatory states by monochromatic driving analogous to that take
place for electronic states of atomic systems. Thus, the limited
two-level and three-level systems can be realized approximately
for oscillatory systems.

An oscillator operated in quantum regime is naturally described by
Fock states that are states with definite numbers of energy
quanta.  The preparation and use of Fock states and various
superpositions of Fock states form the basis of quantum
computation and communications \cite{Niel}. However, excitations
of oscillatory systems usually lead to production of coherent
states nearly indistinguishable from a classical state, but not
quantum Fock states. In this reason quantum oscillatory states are
usually prepared and manipulated by coupling oscillators to atomic
systems. In this way, a classical pulse applied to the atomic
states creates a quantum state that can subsequently be
transferred to the harmonic oscillator excited in a coherent
state.  The systematic procedure has been proposed in \cite{law}
and has been demonstrated for deterministic preparation of
mechanical oscillatory Fock states with trapped ions \cite{mek}
and in cavity QED with Rydberg atoms \cite{varc}. Most recently
the analogous procedure has been applied in solid-state circuit
QED for deterministic preparation of photon number states in a
resonator by interposing a highly nonlinear Josephson phase qubit
between a superconductive resonator \cite{hof}.

Recently, it has been  shown that production of Fock states and
Fock states superpositions or qubits can also be realized  in over-transient regime of an
anharmonic dissipative oscillator
without any interactions with atomic and spin-1/2 systems
\cite{arxiv}.  Preparation of  the lower Fock states $|1\rangle$
and superposition state
$|\Psi\rangle=\frac{1}{\sqrt{2}}(|0\rangle+|1\rangle)$ has been
demonstrated for  the lowest excitation
$|0\rangle\rightarrow|1\rangle$  with complete consideration of
decoherence effects. For this goal  the strong Kerr nonlinearity
as well as the  excitation of resolved lower oscillatory energy
levels with specific train of Gaussian pulses have been
considered.

In continuation of this paper \cite{arxiv} here we propose the
other oscillatory scheme based on two-quantum resolved excitations
of oscillatory levels $|n\rangle\rightarrow|n+2\rangle$.  Thus,
this scheme is the Duffing oscillator for the mode created in the
process of  degenerate down-conversion under an external field.
One of  the possible realizations of this scheme is the
parametrically driven anharmonic oscillator (PDAO) that consists
from intracavity degenerate parametric oscillator combined with
phase modulation element inserted in a cavity.  In this scheme
intracavity oscillatory mode is excited by strong field through
degenerate down-conversion and phase modulation leads to
anharmonicity of oscillatory mode.

Note, that parametrically driven anharmonic oscillator based on
cascaded parametric oscillator and third-order phase modulation
has been proposed and studied in the papers \cite{PNO1},
\cite{PNO2} for the  special cases  without consideration of
effects of dissipation and quantum fluctuations. Quantum theory of
monochromatically driven parametric oscillator combined by phase
modulation  has been developed in terms of the Fokker-Planck
equation in complex P representation \cite{a33}, \cite{b33}. The
exactly solution of this equation (that means consideration of all
order of dissipation) has been obtained in steady-state regime and
the  Wigner function of oscillatory mode has been obtained in
analytical form using this solution \cite{a33}. These results
strongly demonstrate the vanishing of Fock states and quantum
superposition states  in over transient dissipative operational
regime of PDAO within framework of the Wigner function that
visualizes quantum effects as negative values in phase-space.
Really,  analytical results for the Wigner functions of PDAO under
monochromatic excitation \cite{a33} has been calculated as
positive in all ranges of phase space in the steady-state regime.

In this paper, we demonstrate that in the specific pulsed regime
of PDAO the production of nonclassical oscillatory states with
two-fold symmetry in phase-space consisting form the Fock states
$|0\rangle$, $|2\rangle$, as well as superposition of Fock states
$|\Psi\rangle=\frac{1}{\sqrt{2}}(|0\rangle+|2\rangle)$ are
realized. These states can be created for time intervals exceeding
the characteristic time of decoherence.

The quantum regime of PDAO requires a high third-order, Kerr
nonlinearity with respect to dissipation. The largest Kerr
nonlinearities for oscillatory systems were proposed for cooling
nano-electromechanical systems  and nano-opto-mechanical systems
based on various oscillators \cite{craig},\cite{ekinci}.
Superconducting devices based on the nonlinearity of the Josephson
junction (JJ) that exhibits a wide variety quantum phenomena
\cite{21}-\cite{Hosk} offer an unprecedented high level of
nonlinearity and low quantum noise. In some of these devices
dynamics are analogous to those of a quantum particle in an
oscillatory anharmonic potential \cite{claud}, \cite{vijay}. Note,
that comparison of third-order nonlinearities taking place for
various quantum devices have been recently analyzed in
\cite{practice}.

The paper is arranged as follows. In Sec. II we describe
parametrically driven nonlinear oscillator under  pulsed
excitation and descibe phase-space symmetry properties  of the model. In Sec. III
we shortly discuss PDAO under a monochromatic driving for both
transient and steady-state regimes. In Sec.IV we consider
production of nonclassical oscillatory  states in the pulsed regime of PDAO. We
summarize our results in Sec. V.

\section{Model description}

In this section we give the theoretical description of the
system. The nonlinear oscillator driven parametrically by train of pulses and interacting with a reservoir
is described by the following Hamiltonian

\begin{eqnarray}
H=\hbar \omega_{0}a^{+}a + \hbar \chi (a^{+}a)^{2}+
~~~~~~~~~~~~~~~~~~~~~~~\nonumber ~\\ \hbar \Omega (E(t)e^{-i\omega
t}a^{+2} + E^{+}(t)e^{i\omega t}a^{2})+H_{loss}, \label{H}
\end{eqnarray}
where  $a^{+}$, $a$ are the oscillatory creation and annihilation
operators, $ \omega_{0}$ is the oscillatory frequency, $\chi$ is
the nonlinearity strength proportional to the third-order
susceptibility. The coupling constant $\Omega $  is proportional
to the second-order susceptibility and the time-dependent
amplitude of the driving field $E(t)=E_{0}f(t)$  consists from the
Gaussian pulses with the duration $T$ which are separated by time
intervals $\tau$
\begin{equation}
f(t)=\sum{e^{-(t - t_{0} - n\tau)^{2}/T^{2}}}. \label{driving}
\end{equation}
$H_{loss}=a \Gamma^{+} + a^{+} \Gamma$ is responsible for the
linear losses of oscillatory state, due to couplings with heat
reservoir operators giving rise to the damping rate $\gamma$. The
reduced density operator $\rho$ within the framework of the
rotating-wave approximation, in the interaction picture
corresponding to the transformation $\rho \rightarrow e^{-i(\omega
/2)a^{+}at} \rho e^{i(\omega /2)a^{+}at}$ is governed by the master
equation

\begin{eqnarray}
\frac{d \rho}{dt} =-\frac{i}{\hbar}[H_{0}+H_{int}, \rho] + ~~~~~~~~~~~~~~~~~~~~~~~\nonumber ~\\
\sum_{i=1,2}\left( L_{i}\rho
L_{i}^{+}-\frac{1}{2}L_{i}^{+}L_{i}\rho-\frac{1}{2}\rho L_{i}^{+}
L_{i}\right)\label{master},
\end{eqnarray}
where $L_{1}=\sqrt{(N+1)\gamma}a$ and $L_{2}=\sqrt{N\gamma}a^+$
are the Lindblad operators, $\gamma$ is a dissipation rate and N
denotes the mean number of quanta of heath bath,

\begin{eqnarray}
H_{0}=\hbar \Delta a^{+}a, ~~~~~~~~~~~~~~~~~~~~~~~~~~~~~~~~~\nonumber ~\\
H_{int}= \hbar \chi (a^{+}a)^{2} + \hbar \Omega( E(t)a^{+2} +
E(t)^{*}a^{2}),\label{hamiltonian4}
\end{eqnarray}
and $\Delta=\omega_{0} -\omega/2$ is the detuning between half
frequency of the driving field $ \omega/2$  and the oscillatory
frequency $ \omega_{0}$. Two last terms in the interaction
Hamiltonian  describe the self-phase modulation (SPM) of the
oscillatory mode and the parametric three-wave interaction between
semi-classical driving field and the oscillatory mode,
respectively.

To study the pure quantum effects we focus on the cases of very low reservoir temperatures which, however, ought to be still
larger than the characteristic temperature $T \gg T_{cr}=\hbar\gamma/k_B$. This
restriction implies that dissipative effects can be described self-consistently
in the frame of the Linblad Eq. (\ref{master}). For clarity, in our numerical
calculation we choose the mean number of reservoir photons $N=0$.

It is evident that the system displays definite symmetry properties in phase-space. Really,  considering the transformations
\begin{equation}
H^{\prime}=U^{-1}HU,~~ \rho^{\prime}=U^{-1}\rho U
\label{Transform}
\end{equation}
with the unitary operator
\begin{equation}
U=exp\left(i \theta a^{+}a \right). \label{TransformOp}
\end{equation}
we verify that the interaction Hamiltonian (\ref{hamiltonian4})
satisfies the commutation relation
\begin{equation}
\left[H,U\right]=0. \label{ComRel}
\end{equation}
The analogous symmetry takes place for the
density operator of oscillatory mode
\begin{equation}
\left[\rho(t),U\right]=0. \label{DisEf}
\end{equation}

One of the most important conclusions of such symmetries is
related to the Wigner functions of the oscillatory mode
\begin{equation}
W\left(\alpha\right)=\frac{1}{\pi^{2}}\int d^{2}\gamma
Tr\left( \rho e^{\gamma a^{+}-\gamma^{*}a}\right)
e^{\gamma^{*}\alpha -\gamma \alpha^{*}}, \label{Wig}
\end{equation}
where we perform rotations by the angle $\theta$ around the origin in
phase spaces of complex variables $\alpha$ corresponding to
the field operators $a$ in the positive P-representation.
Indeed, in the polar coordinates $r,~\theta$ of the complex
phase-space plane $X=\left(\alpha + \alpha^{*}\right)/2=r\cos
\theta$, $Y=\left(\alpha - \alpha^{*}\right)/2i=r\sin \theta$ we
derive that the Wigner function displays two-fold symmetry in its rotation  around the origin of phase-space
\begin{equation}
W\left(r,\theta+\pi\right)=W\left(r,\theta
\right). \label{WigPol}
\end{equation}
It is well known that this symmetry take place for the degenerate
optical parametric oscillator (OPO) and  reflects on the
phase-locking phenomenon in above threshold regime of OPO.
According to phase-locking  the mode of sub-harmonic generated in
OPO are produced with well-defined two phases \cite{lock}.  The
result (10) shows that this situation takes place also for the
combined system under consideration. The illustrations will be
presented below on the Figs. \ref{li3}, \ref{li5}.

In the following the distribution of oscillatory excitation states
$P(n)=\langle n|\rho|n\rangle$ as well as the Wigner functions
\begin{equation}
W(r, \theta)=\sum_{n,m}\rho_{nm}(t)W_{mn}(r,\theta)
\label{expr:wigner}
\end{equation}
in terms of the matrix elements $\rho_{nm}=\langle
n|\rho|m\rangle$ of the density operator in the Fock state
representation will be calculated. Here  the coefficients
$W_{mn}(r,\theta)$ are the Fourier transform of matrix elements of
the Wigner characteristic function.

In this paper, we demonstrate that in the specific pulsed regime
of PDAO and for the case of strong nonlinearities the Fock states as
well as superposition of Fock states are realized . These states
can be created for time intervals exceeding the characteristic
time of decoherence. The corresponding Wigner functions of
oscillatory mode show ranges of negative values and gradually
deviate from the Wigner function of an PDAO driven by
monochromatic driving.

Nevertheless, the time evolution of PDAO driven by a coherent
force cannot be solved analytically for arbitrary evolution times
and suitable numerical methods have to be used. We solve the
master equation Eq. (\ref{master}) numerically based on quantum
state diffusion method \cite{qsd}. The applications of this method
for studies of the  driven nonlinear oscillators and the
parametric optical oscillator can be found in
\cite{AMK}-\cite{mpop}. In the calculations a finite bases of
number states $|n\rangle$ is kept large enough (with $n_{max}$ is
typically 50) so that the highest energy states are never
populated appreciably.

It should be also noted that the investigation of quantum dynamics
of a driven dissipative nonlinear oscillator for nonstationary
cases, particularly, for various pulsed regimes, is much more
complicated and only a few papers have been done in this field up
to now. Quantum effects in nonlinear dissipative oscillator  with
time-modulated driving force have been studied in the series of
the papers  \cite{AMK}-\cite{qsch}  in the context of a quantum
stochastic resonance \cite{AMK}, quantum dissipative chaos
\cite{K2}, \cite{mpop}, and quantum interference assisted by a
bistability \cite{qsch}.

\section{PDAO in monochromatic field}

In this section, we shortly discuss the  combined dissipative
oscillator under monochromatic excitation, considering $f(t)=1$ in
the Hamiltonian (\ref{hamiltonian4}). At first, we  present the
results in  the semiclassical approach and the standard linear
stability analysis with respect to small deviations from steady
state in terms of the amplitude of the oscillatory mode $\alpha =
n^{1/2}exp(i \varphi)$ \cite{a33}. The intensity $n$ (in photon
number units) and the phase of the mode $\varphi$ in stable
above-threshold regime is determined by the following expressions:

\begin{eqnarray}
n=\frac{\gamma}{2 \chi} [\frac{\Delta}{\gamma}+(J -1)^{1/2}], \nonumber ~\\
\sin(\Phi -2\varphi)=J ^{-1/2},\label{hamiltonian}
\end{eqnarray}
where $J = \frac{ \Omega ^{2}}{\gamma^{2}}I$ and $\Phi$ is the
phase of the driving field $E_{0}=I^{1/2}exp(i\Phi)$. Above
threshold regime takes place for $I >
I_{th}=\frac{\gamma^{2}}{\Omega^{2}}(1+\frac{\Delta^{2}}{\gamma^{2}})$
and the regular behaviour is realized for negative detunings.
These results are obtained in over transient regime and for large
oscillatory mean excitation numbers $n>>1$.

\begin{figure}
\includegraphics[width=8.6cm]{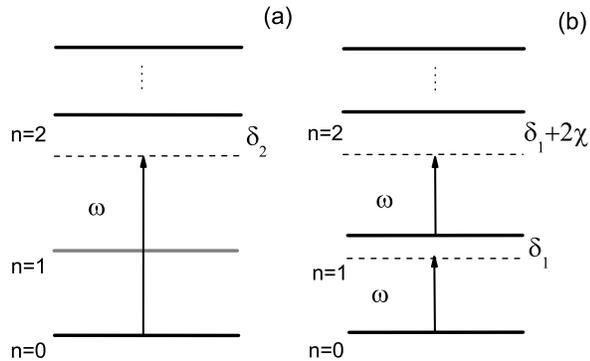}
\caption {Transitions between energetic levels of anharmonic
oscillator: parametric driving with  two-quanta  detunings
$\Delta=\omega_{0} -\omega/2$ and
$\delta_2=\frac{1}{\hbar}(E_2-E_0)-\omega=2\Delta + 4\chi$  (a);
standard driving with one-quanta detunigs $\Delta=\omega_{0}
-\omega$ and $\delta_1=\frac{1}{\hbar}(E_1-E_0)-\omega=\Delta +
\chi$  (b). } \label{li0}
\end{figure}

We now present results for strong  quantum regime of PDNO  that is
realized for $\chi/\gamma \geq 1$. In the absence of any driving,
the quantized vibration states of nonlinear oscillator are the
Fock states $|n\rangle$ which are spaced in energy $E_{n} = E_{0}
+ \hbar\omega_{0} n + \hbar\chi n^2$ with $n = 0, 1, ...$. The
levels form a anharmonic ladder (see, Fig. \ref{li0}(a)) with
anharmonicity that is given by $E_{21}-E_{10}=2\hbar\chi$. Below
we concentrate on quantum regimes for the parameters when
oscillatory energy levels are well resolved considering near to
resonant two-photon transitions between lower number states
$|0\rangle\rightarrow|2\rangle$,  $|1\rangle\rightarrow|3\rangle$.
Thus, we assume non strong excitation regime $E\Omega/\gamma=7$
and that the detuning
$\delta_2=\frac{1}{\hbar}(E_2-E_0)-\omega=2\Delta + 4\chi$  meets
the near to resonant condition, $\delta_2=16\gamma$.  The detuning
between the transition $|1\rangle\rightarrow|3\rangle$ reads as
$\delta_3=\frac{1}{\hbar}(E_3-E_1)-\omega=2\Delta + 8\chi$ and for
this parameters is more large than $\delta_2$, i.e.
$\delta_3=36\gamma$. For comparison we add in Fig. \ref{li0}(b)
also the lower levels and the oscillatory transitions for the
scheme considered in \cite{arxiv}.

\begin{figure}
\includegraphics[width=8.6cm]{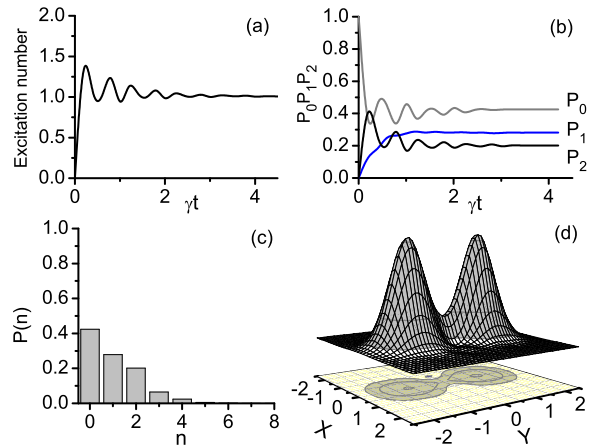}
\caption{Time evolution of the averaged excitation numbers (a); the Rabi oscillations of the
state populations with decoherence which suppresses beating (b);
the distribution of excitation numbers  (c); and the Wigner function
(d) for PDAO in steady-state regime. The parameters are: $\Delta/\gamma=
-2$, $\chi /\gamma = 5$, $E\Omega/\gamma = 7$} \label{li1}
\end{figure}

\begin{figure}
\includegraphics[width=8.6cm]{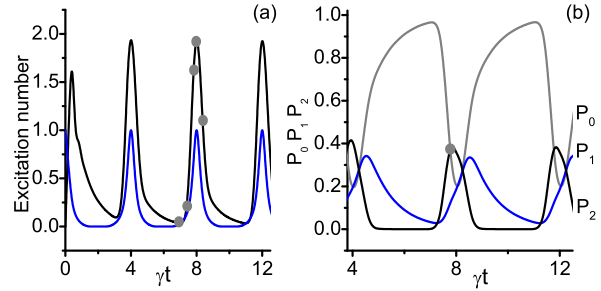}
\caption{Time-evolution of averaged excitation number (a) and
populations of Fock states (b). Time-dependent structure of pulses
is indicated below n(t) in arbitrary units. The parameters are:
$\Delta/\gamma= -2$, $\chi /\gamma = 5$, $E\Omega/\gamma = 10$,
$\tau = 4{\gamma}^{-1}$, $T=0.5{\gamma}^{-1}$} \label{li2}
\end{figure}

Now, we present  on Fig. \ref{li1} the numerical results for
monochromatically driven PDAO in dependence of the parameters:
$\chi, \Delta, E\Omega$.  Transition the system to the steady
state for time intervals $t\gg\gamma^{-1}$ is depicted in Figs.
\ref{li1} (a,b) on  the time-dependent mean excitation number and
the populations of Fock states.  As we noted above,  steady state
results, in that number probability distribution of oscillatory
excitation numbers and the Wigner function of oscillatory mode
have been obtained analytically in terms of the exact solution of
the Fokker-Planck equation \cite{a33}, \cite{b33}. Particularly,
the solution for the Wigner function of oscillatory mode involving
quantum noise in all order of perturbation theory and in
steady-state regime  is positive in all phase space and hence does
not describe  Fock states in over-transient regime. Note, that the
steady-state solution of the Fokker-Planck equation has been found
using the approximation method of potential equations
\cite{Drummond}. The validity of this solution has not been
checked in the strong quantum regime of PDAO operated on the level
of few excitation number. On this reason we calculate the Wigner
function numerically on the base of numerical simulation of master
equation by using quantum state diffusion method \cite{qsd}.  The
results are  displayed in the Figs.\ref{li1} (d). As we see, the
Wigner function displays two humps and has two-fold  symmetry  in
phase-space under the rotation on angle $\pi$  around its origin
(10). This effect reflects the well-known phenomena of phase
locking that takes place for both OPO and OPO combined with phase
modulation. In semiclassical approach according to the equation
(12) phase-locking in above threshold regime means the forming of
two stable states of oscillatory mode with equal photon numbers,
but with two  different phases which are $\Phi/2$ and
$\Phi/2+\pi$. In quantum treatment of PDAO this phenomenon is
displayed as two-humped structure of the Wigner function as is has
been demonstrated analytically in \cite{b33}. Here we demonstrate
that similar effect occurs also for PDAO  in strong quantum regime
on the level of few excitation numbers.

\section{Production of oscillatory quantum states in the pulsed regime}

Now we are able to present the results concerning production of
oscillatory quantum states  for pulsed excitation regime of PDAO.
Since in this paper we focus on a system that remains close to its
ground, vacuum state, we assume a small number of excitation
number in the oscillator. We consider interaction time intervals
exceeding the characteristic time of dissipative processes,
$t\gg\gamma^{-1}$, however, in the nonstationary regime that is
conditioned by the specific form of pulsed excitation. If the
parametric excitation stipulated by train of the Gaussian pulses
the ensemble-averaged mean oscillatory excitation numbers, the
populations of oscillatory states and the Wigner functions are
nonstationary and exhibit a periodic time dependent behavior, i.e.
repeat the periodicity of the laser pulses in an over transient
regime.

\begin{figure*}
\begin{center}
\includegraphics[width=18cm]{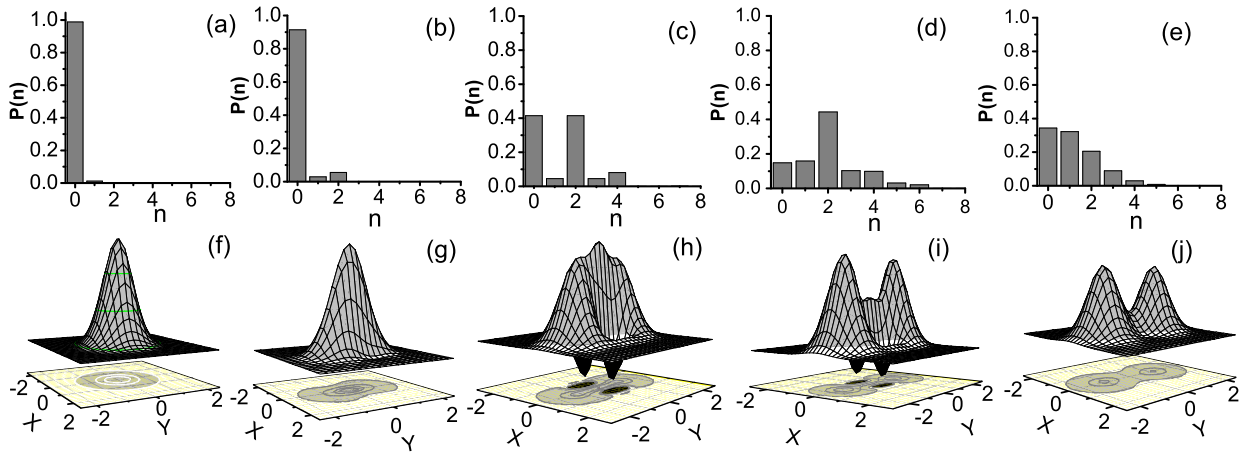}
\end{center}
\caption{The excitation number distributions for definite
time-intervals (gray points noted in Figs. 3(a)):  $t=2\tau-2T $
(a); $t=2\tau-1.8T $ (b); $t=2\tau-0.4T $ (c); $t=2\tau $ (d);
$t=2\tau+0.6T $ (e). The Wigner functions corresponding these
distributions for the same time- intervals. The black circles
indicate the negative parts of the Wigner functions. The
parameters are: $\Delta/\gamma= -2$, $\chi /\gamma = 5$, $E
\Omega/\gamma = 10$, $\tau = 4{\gamma}^{-1}$,
$T=0.5{\gamma}^{-1}$.} \label{li3}
\end{figure*}

\begin{figure}
\includegraphics[width=8.6cm]{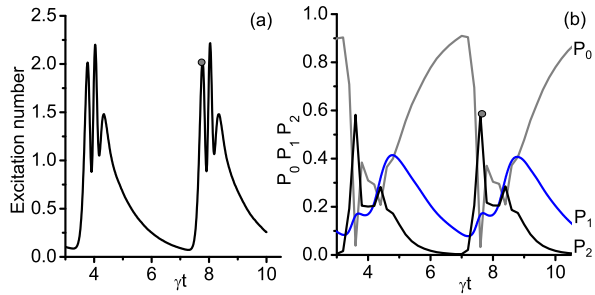}
\caption{Time evolution of averaged excitation number (a) and
populations of Fock states (b). The parameters are:
$\Delta/\gamma= -10$, $\chi /\gamma = 5$, $E \Omega/\gamma =
10.3$, $\tau = 4{\gamma}^{-1}$, $T=0.5{\gamma}^{-1}$} \label{li4}
\end{figure}

\begin{figure}
\includegraphics[width=8.6cm]{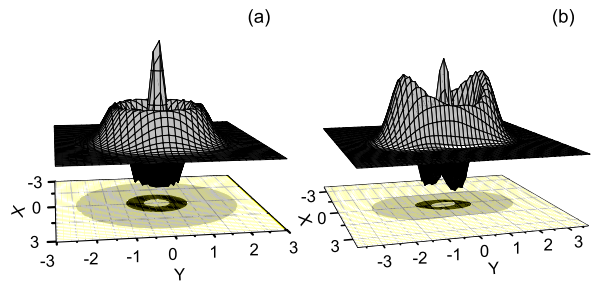}
\caption{The Wigner function for pure $|2\rangle$ state (a). The
Wigner function for $t=2\tau - 0.8T$ (b), for the parameters:
$\Delta/\gamma=-10$, $\chi /\gamma = 5$, $E \Omega/\gamma = 10.3$,
$T= 0.5{\gamma}^{-1}$, $\tau =4{\gamma}^{-1}$. The ranges of
negativity are indicated in the black.} \label{li5}
\end{figure}

As it has been shown  in  \cite{arxiv} the most striking signature
of periodically pulsed NDO is the appearance of various quantum
states involving Fock states and interference between Fock states
in over transient regime. Excitation due to parametric interaction
leads to production of new class of oscillatory states that have
not been possible to generate in standard way by one-photon
excitation.  These states include squeezed states or
superpositions of Fock state, as well as quantum localized states
on the level of few excitation numbers. We demonstrate this point
for near to two-quanta resonant transitions
$|n\rangle\rightarrow|n+2\rangle$ for the oscillatory parameters
used above in Figs. \ref{li2} and \ref{li3} however, for the
periodically  pulsed regime with the  duration $T =
0.5\gamma^{-1}$, and the time interval between pulses $\tau =
4\gamma^{-1}$. The evolution of the averaged excitation numbers
and the population of the lowest states $|0\rangle$, $|1\rangle$
and $|2\rangle$  are depicted in Figs. \ref{li2} while
time-dependent structure of pulses is indicated below $n(t)$ in
arbitrary units. As we see, for over transient regime the
time-modulation of the averaged excitation number and the
populations of oscillatory states repeat the periodicity of the
pump laser. Besides these results, the excitation numbers for the
definite time-intervals during pulses and the corresponding Wigner
functions of oscillatory mode are shown in Figs. \ref{li3}. As we
see the evolution of anharmonic oscillator under parametric
excitation involves number of quantum oscillatory states which are
visualized on the Wigner functions for typical five measurement
time-intervals. We indicate the choosing of these time-intervals
on the curve of main excitation number (see, Fig.\ref{li2} (a)).
For $t=k\tau-2T$, k = 2, 3, ... the oscillatory mode is in the
vacuum state,therefore the corresponding Wigner function is
Gaussian (see,  Fig.\ref{li3} (f)). After evolution of the system,
for the time intervals $t=k\tau-1.8T$, the state is squeezed in
phase space. The typical result characterizing parametric double
excitation  of oscillatory mode is effective simultaneous
production of  $|0\rangle$ and $|2\rangle$ states that are
depicted in Fig. \ref{li3} (c) for the time intervals
$t=k\tau-0.4T$. The populations of these states is approximately
equal one to the other and the corresponding Wigner function (see,
Fig.\ref{li3} (h)) demonstrates the interference fringes on
phase-space between $|0\rangle$ and $|2\rangle$ states. It is easy
understand that this Wigner function  approximately describes the
pure superposition state
$|\Psi\rangle=\frac{1}{\sqrt{2}}(|0\rangle+|2\rangle)$, i.e. for
these time-interval PDAO produces superposition of Fock states
$|0\rangle$ and $|2\rangle$. It is important that this quantum
interference effect  is realized when the oscillatory mean
excitation number reaches its maximal value.  For further
increasing of time interval, i.e. for  $t=k\tau$,  the
interference fringes on phase-space are deformed and the Wigner
function consists from two localized peaks with the ranges of
negativity between them (see, Fig.\ref{li3} (d, i)). Then, near to
the end of the pulses, i.e. $t=k\tau+0.6T$, the quantum
interference is vanished and production of localized state takes
place for PDAO (see,  Fig.\ref{li3} (e,f)) as has been obtained in
the steady state regime. It should be noted that these results are
in accordance with the phase symmetry properties of PDAO (10). We
conclude that production of quantum interference is realized in
the vicinity of time-intervals where populations of $|0\rangle$
and $|2\rangle$ states are crossed (see, Fig \ref{li2}(b)) in over
transient regime. We assume that the control of decoherence in
this case take place due to application of suitable tailored,
synchronized pulses (see, for example, \cite{kik}, \cite{ent}).
Indeed, quantum interference is realized here if a mutual
influence of pulses is essential, (for $\tau/T = 3.14$ on Fig.
\ref{li3}).

The next regime that we consider is that where the pulsed
excitation is tuned to the exact resonance, $\delta=2\Delta +
4\chi=0$. It is evident that in this case the Fock state
$|2\rangle$ can be effectively produced if a low excitation is
used. Similarly to what was done above in dispersive regime, we
consider at first the evolution of averaged excitation numbers as
well as  the population of the lowest states $|0\rangle$,
$|1\rangle$ and $|2\rangle$ by two-quanta excitation. The results
depicted in Figs. \ref{li4} allows us to choose the time-intervals
within a pulse for which the maximal probability of production of
$|2\rangle$ Fock state is realized. The time evolution of the
probabilities $P_0$,  $P_1$  and  $P_2$ of $|0\rangle$,
$|1\rangle$ and $|2\rangle$ states starting from the vacuum state
are depicted in Fig. \ref{li4}(b). As we see, that the maximal
weight 0.6 of $P_2$ is realized for the definite time-intervals of
measurement $t = k\tau - 0.25T$, $k = 1, 2, 3 ...$, (see, Fig.
\ref{li4} (a)). In this way, for the duration of pulses  $T =
0.5\gamma^{-1}$, and the time interval between them $\tau =
4\gamma^{-1}$  the Wigner function is depicted in Fig.
\ref{li5}(b) for time measurement intervals $t=k\tau+0.8T$. One
might expect that this result would be approximately close to the
pure $|2\rangle$ Fock state Wigner function displayed for
comparison in Fig. \ref{li5}(a). This Wigner function displays
ring signature with the center at $x = y = 0$ in phase-space.
Indeed, in the general these results are qualitatively similar
involving also negative part, however, the cyclic symmetry of pure
$|2\rangle$ has not displayed in state of PDAO that acquires
two-fold symmetry in phase-space due to parametric excitation.

\section{Conclusion}

We have  demonstrated  production of various quantum  states for
the parametrically driven anharmonic oscillator in the regimes of
low excitation and in complete consideration of dissipative
effects. This investigation continues our previous analysis
\cite{arxiv} devoted to creation of Fock states as well as
superpositions of Fock states in the specific regime of
periodically pulsed anharmonic oscillator  for time-intervals
exceeding the characteristic decoherence time. In this paper, we
have proposed the other nonlinear oscillatory system that is
excited by the train of Gaussian laser pulses through the
degenerate down-conversion process. It can be realized as OPO
combined with phase modulation. Preparation of quantum states in
PDAO has been stipulated by strong Kerr nonlinearity as well as by
two-quanta resonant condition. We have studied the role of
phase-localizing processes on production of oscillatory
nonclassical states on the level of few excitation numbers. In
this way, the production of the states approximately close to  the
superposition state
$|\Psi\rangle=\frac{1}{\sqrt{2}}(|0\rangle+|2\rangle)$ and
$|2\rangle$  Fock state have been described.

\begin{acknowledgments}
We  acknowledge helpful discussions with Chew Lock Yue.
\end{acknowledgments}

\end{document}